\documentclass[pra,twocolumn,showpacs,floatfix]{revtex4-1}

\usepackage[T1]{fontenc}
\usepackage[utf8]{inputenc}
\usepackage{bm,bbm}
\usepackage{graphicx}
\usepackage{color}

\usepackage{hyperref}
\usepackage[ngerman,english]{babel}
\usepackage{amsmath,amssymb,amsfonts}
\usepackage{dcolumn}
\usepackage{float}
\usepackage{mathrsfs}
\usepackage[cmyk]{xcolor}

\newcommand{\Real}{\mathrm{Re}\,}
\newcommand{\Imag}{\mathrm{Im}\,}
\newcommand{\abs}[1]{\left| #1 \right|} 					
\newcommand{\ii}{\mathrm{i}}							
\newcommand{\ee}[1]{\mathrm{e}^{#1}}						
\newcommand{\dd}{\mathrm{d}}							
\renewcommand{\vec}[1]{\bm{#1}}							
\newcommand{\vectornorm}[1]{\left|\left| #1 \right|\right|}			

\newcommand{\del}[2]{\frac{\partial #1}{\partial #2}}				
\newcommand{\PT}{\ensuremath{\mathcal{PT}~}}
\newcommand{\PTs}{\ensuremath{\mathcal{PT}\!}-}                   
\newcommand{\sub}[1]{_{\mathrm{ #1 }}}                          

\newcommand{\etal}{et al.\ }

\newcommand{\Eq}{Eq.}
\newcommand{\Sec}{Sec.}
\newcommand{\Fig}{Fig.}
\newcommand{\Figs}{Figs.}
\newcommand{\Ref}{Ref.}
\newcommand{\Refs}{Refs.}

\begin{document}

\title{Dipolar Bose-Einstein condensates in a \PTs symmetric
  double-well potential}

\author{R\"udiger Fortanier}
\email{ruediger.fortanier@itp1.uni-stuttgart.de}
\author{Dennis Dast}
\author{Daniel Haag}
\author{Holger Cartarius}
\author{J\"org Main}
\author{G\"unter Wunner} 
\author{Robin Gut\"ohrlein} 
\affiliation{Institut f\"ur Theoretische Physik 1, Universit\"at Stuttgart,
70550 Stuttgart, Germany}

\date{\today}

\begin{abstract}
  We investigate dipolar Bose-Einstein condensates in a complex
  external double-well potential that features a combined parity and
  time-reversal symmetry. On the basis of the Gross-Pitaevskii
  equation we study the effects of the long-ranged anisotropic
  dipole-dipole interaction on ground and excited states by the use of
  a time-dependent variational approach. We show that the property of
  a similar non-dipolar condensate to possess real energy eigenvalues
  in certain parameter ranges is preserved despite the inclusion of
  this nonlinear interaction. Furthermore, we present states that
  break the \PT symmetry and investigate the stability of the distinct
  stationary solutions. In our dynamical simulations we reveal a
  complex stabilization mechanism for \PTs symmetric, as well as for
  \PTs broken states which are, in principle, unstable with respect to
  small perturbations.
\end{abstract}


\maketitle

\section{Introduction}
\label{sec:introduction}

Open quantum systems and their effects, such as damping, dephasing,
resonance phenomena and more, can be described by non-Hermitian
Hamiltonians \cite{MoiseyevBook2011}. In particular, there are
problems where the non-Hermitian quantum mechanics formalism is
necessary, as e.g.\ in quantum field theory, or where it supports a
simple description as in optics for a complex refraction index, or in
cases where complex potentials are introduced \cite{MoiseyevBook2011}.
Furthermore, it is sometimes quite advantageous to use non-Hermitian
Hamiltonians although the problem could, in principle, be solved
within the conventional Hermitian framework. One of the most important
applications is the modeling of dissipation and influx by a complex
potential. The formalism has been originally developed for the linear
Schrödinger equation, however, it is applicable to the nonlinear
Gross-Pitaevskii equation (GPE), as well. A widely-used procedure is
the implementation of inelastic three-body losses in terms of an
imaginary potential \cite{Koeberle2012}. Moreover, non-Hermitian forms
of the GPE provide access to the decay of the condensates with the
complex scaling approach \cite{Moiseyev05,Schlagheck06}, to transport
phenomena \cite{Paul07}, and to the theoretical study of dissipative
optical lattices \cite{Abdullaev10,Bludov10}. An analytical
continuation of the GPE made it possible to discover exceptional
points (EPs) and study their properties
\cite{Cartarius08a,Gutoehrlein13}.

One special class of non-Hermitian Hamiltonians are the \PTs symmetric
ones. They commute with the \PT operator, which combines the action of
parity and time reflection, i.e.\ $\left[\mathcal{PT},H\right]=0$.
Bender and Boettcher \cite{Bender1998,Bender1999} found that real
eigenvalues are possible despite the non-Hermiticity of those
Hamiltonians and in a certain parameter range completely real
eigenvalue spectra can exist . Experimental observations of \PT
symmetry have been achieved in optical systems
\cite{Guo09,Rueter2010,Regensburger2012}, yet no \PTs symmetric
genuine quantum system could be realized. In
\Refs~\cite{Cartarius2012,Dast2013} it has been shown that
Bose-Einstein condensates (BECs) in \PTs symmetric double-delta and
double-well potentials, respectively, constitute such systems.
Furthermore, in \Refs~\cite{Dast2013,Dast2013jphysa} the effect of the
short-ranged contact interaction of the particles on the stationary
states and on the dynamics has been investigated. There, good
agreement with the results obtained by Graefe \etal
\cite{Graefe2008JPhysA,Graefe2010} via a simple matrix model has been
proven.

The experimental realization of atoms sustaining a large magnetic
dipole moment e.g.\ ${}^{52}\mathrm{Cr}$
\cite{Griesmaier05a,Beaufils2008,Lahaye09} and, more recently,
$^{164}$Dy \cite{Lu2010,Lu2011} and ${}^{168}\mathrm{Er}$
\cite{Aikawa_2012} as well as the fast progress towards the creation
of BECs of polar molecules \cite{Ni_2008}, which possess large
electric dipole moments, opened the field of research for effects
generated by the dipole-dipole interaction (DDI). It has been shown in
\Ref~\cite{Dast2013jphysa} by mathematical arguments concerning the
DDI that effects, typical for \PTs symmetric systems, are expected to
be present in a dipolar \PTs symmetric system, as well. Considering
not only the short-ranged contact, but also the long-ranged
anisotropic DDI leads to interesting questions for \PTs symmetric
systems: What is the impact of the DDI on the stationary states? In
particular, real eigenvalues require \PTs symmetric wave functions,
for which the effects of gain and loss modeled by imaginary potentials
are balanced. This relation between true stationary states with real
eigenvalues and the symmetry of the wave function also holds in the
nonlinear GPE
\cite{Cartarius2012,Dast2013,Dast2013jphysa,Graefe2008JPhysA,Graefe2010}
and is even more important since a wave function breaking the \PT
symmetry of the linear potential may even destroy that of the total
nonlinear Hamiltonian. It is well known that the anisotropic DDI can
lead even to structured ground states not possessing the potential's
symmetry. In combination with gain and loss effects this opens the
door for new cases of \PT symmetry breaking. The scenario for BECs in
a \PTs symmetric potential that feature solely short-ranged
interactions can be described by a simple matrix model, as shown by
Graefe \etal \cite{Graefe2008JPhysA,Graefe2010} for a two-mode
Bose-Hubbard system. These results are in agreement with a mean-field
description within the GPE \cite{Dast2013jphysa}. We expect new
effects to arise from the DDI and particularly from its long-range
nature that is not included in the matrix model. Furthermore, dipolar
condensates often feature novel dynamical properties and provide
effects such as long-ranged Josephson oscillations
\cite{Asad09,Xiong09a} or pattern formation \cite{Nath2009} and it is
therefore interesting to explore the time evolution of dipolar
condensates in a \PTs symmetric potential.

\begin{figure}[t]
  \centering
    \includegraphics[width=\columnwidth]{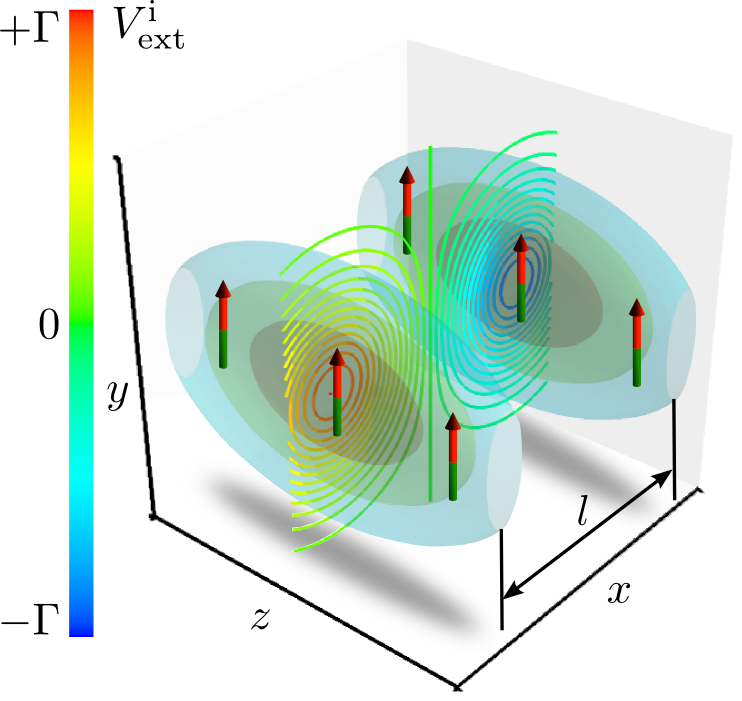}
    \caption{(Color online) Visualization of the \PTs symmetric
      double-well potential in the repulsive configuration. The arrows
      indicate the direction of the dipoles. Isosurfaces show the
      shape of the real part $V_{\mathrm{ext}}^\mathrm{r}$ of the
      external potential, where $l$ is the inter-well spacing as given
      in the text. The contour lines show a slice cut through the
      center of the imaginary part $V_{\mathrm{ext}}^\ii$ of the
      external potential, with the corresponding colorbars on the
      left.}
    \label{fig:PT-geometry}
\end{figure}

In this work we combine the ingredients of the long-ranged DDI with
the \PT symmetry of an external potential. Thereby we consider a
dipolar condensate in a \PTs symmetric double-well potential and
extend the work presented in \Ref~\cite{Dast2013} to dipolar BECs. Our
analysis is based on the time-dependent extended Gross-Pitaevskii
equation
\begin{eqnarray}
  \label{eq:GPE-multi-well-units}
  \ii \frac{\dd}{\dd t} \Psi \left( \vec r, t \right) = 
  \left[
  - \frac{1}{2}\Delta
  + V_{\mathrm{ext}}  + V_{\mathrm{c}}  + V_{\mathrm{dd}}
  \right]
  \Psi \left( \vec r, t \right) \,,
\end{eqnarray}
where $V_{\mathrm{ext}}$, $V_{\mathrm{c}}$, and $V_{\mathrm{dd}}$
denote the external, short-ranged contact and long-ranged
dipole-dipole interaction potential, respectively. In
\Eq~(\ref{eq:GPE-multi-well-units}) and for the potentials below, we
choose the units of energy $E$, time $t$, and length $\vec r$ such
that $[E]=\hbar^2/(ml^2)$, $[t]=ml^2/\hbar$, and $[\vec r]=l$, where
$l$ is the distance between the centers of the two wells (cf.\
\Fig~\ref{fig:PT-geometry}).
The external potential is modeled in our calculations by
\begin{eqnarray}
  \label{eq:PT-potential}
  V\sub{ext} (\vec r) &= 
  -\left( V_0 -\ii \Gamma \right)g^+
  -\left( V_0 +\ii \Gamma \right)g^- \,,\\
\text{with}\quad
  g^\pm  &= \exp \left(
    -\frac{(x \pm
      l/2)^2}{2L_x^2}-\frac{y^2}{2L_y^2}-
    \frac{z^2}{2L_z^2}\right)\,,
\end{eqnarray}
where $V_0$ is the depth of the real double-well potential and
$\Gamma$ is the strength of the gain and loss terms. We choose the
identical parameters as given in \Refs~\cite{Peter12,Fortanier13a} for
a similar triple-well system $V_0=80$, $L_x= L_z = 1/4$, and $L_y =
2$, as these have a reasonable magnitude for a possible corresponding
experiment. The contact interaction potential $V_{\mathrm{c}}$ reads
\begin{eqnarray}
  \label{eq:scattering-potential}
  V_{\mathrm{c}} = 4\pi N a \abs{\Psi \left( \vec r , t \right) }^2 \,,
\end{eqnarray}
with the scattering length $a$ and the number of particles $N$. The
DDI potential is given by
\begin{eqnarray}
  \label{eq:DDI-potential}
  V_{\mathrm{dd}} = 3 N a_{\mathrm{dd}} \int \dd^3 r' 
  \frac{1-3 \cos^2 \theta }{\abs{\vec r - \vec r'}^3} \abs{\Psi
    \left( \vec r', t \right)}^2\,,
\end{eqnarray}
where $a_{\mathrm{dd}}$ is the dipole strength and $\theta$ is the
angle between the vector $\vec r - \vec r^\prime$ and the
direction of the dipole alignment. The dipolar interaction breaks the
symmetry and provides two possible configurations, namely the
repulsive configuration, shown in \Fig~\ref{fig:PT-geometry}, where
the dipoles are aligned in $y$-direction and the attractive
configuration, where the dipoles are aligned in $x$-direction. We will
only discuss the repulsive configuration, yet we have also performed
calculations in the attractive configuration, but found a
qualitatively similar behavior.

\section{Method}
\label{sec:method}

Our method is based on the time-dependent variational principle (TDVP)
with the variational ansatz consisting of a linear superposition of
two Gaussian wave packets (GWPs), $\Psi=g^1+g^2$. Details of the
method can be found in \cite{Eichler12}, where the same method has
been used to describe the collision of quasi-2$d$ anisotropic solitons
and in \cite{Fortanier13a}, where dipolar BECs in triple-well
potentials have been investigated. Yet, we will recapitulate the major
steps for the reader's convenience here. Each of the GWPs has the form
\begin{eqnarray}
  \label{eq:GWP-single-ansatz}
  g^k = \ee{-
      \left(
        \left(
          \vec x - \vec q^k
        \right)^T A^k
        \left(
          \vec x - \vec q^k
        \right) - \ii\left(\vec p^k\right)^T
        \left(
          \vec x - \vec q^k
        \right) +  \gamma^k
      \right)}\,,
\end{eqnarray}
where $k=1,2$; the symbol $T$ denotes the transposition and where in
general the time-dependent parameters $A^k$ are $3\times 3$ complex
diagonal matrices, $\vec p^k$ and $\vec q^k$ are real 3$d$ vectors,
and $\gamma^k$ are complex numbers. We assume that the $y$-direction
(the direction of the dipole alignment) has a strong confinement due
to the external trap and thus ignore translations and rotations in
this direction i.e. $q_y^k \equiv 0$, $A^k_{xy}=A^k_{yz}\equiv 0$.
However, for the other directions we apply no further restrictions,
particularly with respect to position and movement of the GWPs in the
$x$-direction. It is reasonable to start with one GWP placed at the
center of each well.

To determine the time development of the variational parameters we
make use of the TDVP in the formulation of McLachlan
\cite{McLachlan1964a}
\begin{eqnarray}
  \label{eq:McLachlan}
  I = \vectornorm{\mathrm{i} \phi - H \Psi(t)}^{2}\stackrel{!}{=}\min\,,
\end{eqnarray}
where $\phi$ is varied and set $\phi \equiv \dot \Psi$ afterwards. We
then apply the ansatz
\begin{eqnarray}
  \label{eq:ansatz_tdvp_eom}
    \Psi &= g^1 + g^2
\end{eqnarray}
for the variational wave function, which yields the equations of
motion (EOM) for the variational parameters
\begin{eqnarray}
  \label{eq:eom}
  \dot{\vec z} = \vec f\left( \vec z (t) \right) = \vec f\left(A^k (t),\vec q^k (t),\vec
  p^k (t) ,\gamma^k (t) \right)\,,
\end{eqnarray}
with $\vec z = (\vec z^1,\vec z^2)$ and $k=1,2$.

The stationary states of the GPE are the fixed points of
\Eq~\eqref{eq:eom} and can be determined by a nonlinear root search
(e.g.\ Newton-Raphson). An alternative to find the real ground state
is the application of imaginary time evolution (ITE) to the EOM.
However, the ITE does not always converge to the ground state
\cite{Fortanier13a,menotti07a}. To evolve the EOM in imaginary time as
well as in real time, a standard algorithm like Runge-Kutta can be
used.

We investigate the linear stability of the fixed points by the
calculation of the eigenvalues $\Lambda=\Lambda^{\mathrm{r}}+\ii
\Lambda^{\mathrm{i}}$ of the Jacobian
\begin{eqnarray}
  \label{eq:jacobian}
  J= \del{\left( \Real \dot A^k, \Imag \dot A^k, \dot{\vec q}^k,
      \dot{\vec p}^k, \Real \dot{\gamma}^k, \Imag \dot{\gamma}^k \right)}{\left(
      \Real A^j, \Imag A^j, {\vec q^j}, {\vec p^j}, \Real
      {\gamma^j}, \Imag {\gamma^j} \right)}\,,
\end{eqnarray}
with $k,j=1,2$. The eigenvalues $\Lambda$ appear in pairs of opposite
sign and correspond to excitations described by the
Bogoliubov-de~Gennes equations
\cite{Kreibich12a,Kreibich13b,Rau10aI,Rau10aII}. If all real parts
$\Lambda^{\mathrm r}$ vanish, the fixed point is stable, otherwise it is
unstable.

The parameter space is essentially spanned by the three parameters
$Na_\mathrm{dd}$, $Na$, and $\Gamma$, whereas the particle number $N$
is no independent quantity due to the scaling properties of the GPE.
For reasons of clearness we keep the dipole strength constant at $N
a_\mathrm{dd}=0.3$. In order to obtain the states over the range of
interest for the remaining two parameters it is reasonable in the
numerical computation to start at $N a=0$ and $\Gamma=0$, where the algorithm
is most stable and ground states are accessible by an imaginary-time
evolution. Afterwards we advance with the result as initial guess for
the nonlinear root search. However, it turned out that this method
does not guarantee to find the correct states for several reasons. The
number of existing states is not constant and, as we will see later
on, states emerge and vanish in bifurcations. Furthermore, crossings
of states appear. Then, an extrapolation of the previously obtained
variational parameters may lead to a unfortunate choice of initial
parameters, and only a subtle algorithm is capable to advance them. We
use any of the following strategies to obtain results in such cases:
we circumvent the crossings by advancing the other parameter, we
perform forward-jumps and proceed to calculate backwards, or we
perform an extrapolation of the variational parameters close to a
bifurcation point depending on the bifurcation type (e.g.\ square root
behavior of a tangent bifurcation).

\section{Results}
\label{sec:results}

The presence of the dipolar interaction strongly influences the
results for the stationary states in all kinds of real external
potentials. We therefore briefly summarize the picture that one
obtains for the case $\Gamma=0$. A BEC in a real double-well
potential, with no gain or loss present, exhibits spontaneous symmetry
breaking, also known as macroscopic quantum self-trapping, above a
critical value of the scattering length. This effect breaks the
symmetry of the external trap and occurs both in a system with dipolar
and pure contact interactions \cite{Xiong09a,Raghavan99,Schmelcher06}.
To distinguish between the effects originating from short-ranged and
long-ranged interactions in a real external potential it is more
appropriate to choose a triple-well system
\cite{Fortanier13a,Peter12,Zhang12,Lahaye10a}. However, regarding the
stationary states, dipolar effects can be identified as we will show
in \Sec~\ref{sec:stat-stat-dipol}.

\subsection{Stationary states of non-dipolar BECs}
\label{sec:stationary-states}

It turns out that the spectra we obtain for the dipolar condensate in
the \PTs symmetric double well involve more states than in the
non-dipolar case. To understand the influence of the DDI we want to
relate our findings to the simpler case of a BEC which only possesses
short-ranged interactions. In \Ref~\cite{Dast2013} results have been
presented for the non-dipolar system, yet within a different unit
system and with a different specific form of the external potential.
For the convenience of the reader we here qualitatively confirm these
results for the potential (\ref{eq:PT-potential}) with the units given
above and subsequently compare the findings in the dipolar system with
them.

\begin{figure}[tb]
  \centering
  \includegraphics[width=\columnwidth]{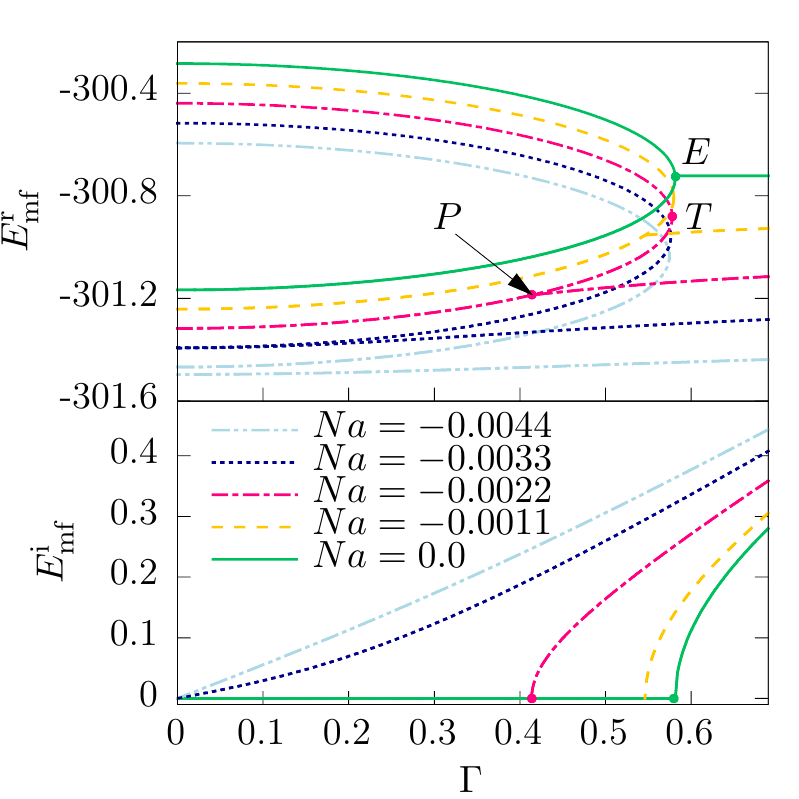}
  \caption{(Color online) Mean-field energy as a function of the
    gain-loss parameter $\Gamma$ at different values of the
    nonlinearity $Na$ for the non-dipolar case. The upper panel shows
    the real and the lower one the imaginary part. For $Na=-0.0022$
    the tangent bifurcation is marked by a dot labeled $T$ and the
    pitchfork bifurcation by a dot labeled $P$. For $Na=0$ both
    bifurcations merge in the singular point $E$. $E_\mathrm{mf}$ and
    $\Gamma$ are given in the units introduced in
    \Sec~\ref{sec:introduction}.}
  \label{fig:non-dipolar}
\end{figure}

In \Fig~\ref{fig:non-dipolar} real and imaginary parts of the
mean-field energy for the non-dipolar case are shown. For vanishing
nonlinearity $Na=0$ two \PTs symmetric states with purely real
eigenvalues exist from $\Gamma =0$ up to a critical value of the
gain-loss parameter $\Gamma$. At this point, labeled $E$ in
\Fig~\ref{fig:non-dipolar}, these states vanish in a bifurcation and
two \PTs broken states with complex conjugate energies emerge. For
these states only $E^{\mathrm{i}}_{\mathrm{mf}} \geq 0$ is shown. As
soon as the nonlinearity is present the tangent bifurcation at point
$E$ splits up into a tangent bifurcation $T$, where the \PTs symmetric
states vanish, and a pitchfork bifurcation $P$, where the \PTs broken
states emerge. Note that although being solutions of the
time-independent GPE, the \PTs broken solutions are no true stationary
solutions for $\Gamma\neq 0$ due to the imaginary part in their energy
eigenvalue. It has been shown in \Refs~\cite{Heiss13,Dast2013jphysa}
that the bifurcation points are exceptional points (EPs) of order 2
and 3. The singular point $E$ has some unique properties as shown in
\Ref~\cite{Dast2013jphysa}. There, an analytic continuation of the GPE
has been performed. Due to the analytic continuation the number of
states does no longer change at the bifurcation points $P$ and $T$.
Two analytically continued states appear in addition to those shown in
\Fig~\ref{fig:non-dipolar} for $\Gamma$ values below $P$. Similarly
two additional analytically continued states are found for $\Gamma$
values above $T$. The extended states always exactly emerge in the
bifurcation points, i.e.\ in the whole range of $\Gamma$ always four
eigenstates are present in the shown energy range. In principle,
excited states with much higher energies $E_\mathrm{mf}^\mathrm{r}$
could be found in the region $V_0 < E_\mathrm{mf}^\mathrm{r} < 0$, yet
these states are energetically separated far enough as to influence
the states investigated here. This argument can be applied to the
dipolar system as well. In the limit $Na\rightarrow 0$ the points $P$
and $T$ merge and it can be shown that structures of a fourth-order EP
are revealed. However, these structures are only observable as long as
$Na\neq 0$. For $Na=0$ the analytic continuation of the GPE splits
into two uncoupled parts that are equivalent to the non-extended GPE.
Thus, only two identical spectra on top of each other with two
identical second-order EPs at the only remaining bifurcation $E$ are
observed.

\subsection{Stationary states of dipolar BECs}
\label{sec:stat-stat-dipol}

We will now include the dipolar interaction and investigate the system
with the results of the non-dipolar case in mind. Although we
performed calculations for a wide range of the parameters $Na$ and
$\Gamma$ we will concentrate on a domain found to express interesting
phenomena. Still, keep in mind that the following effects and
characteristics can be present in different parameter regions for a
different dipole strength $Na_\mathrm{dd}$.
\begin{figure}[tb]
  \centering
  \includegraphics[width=\columnwidth]{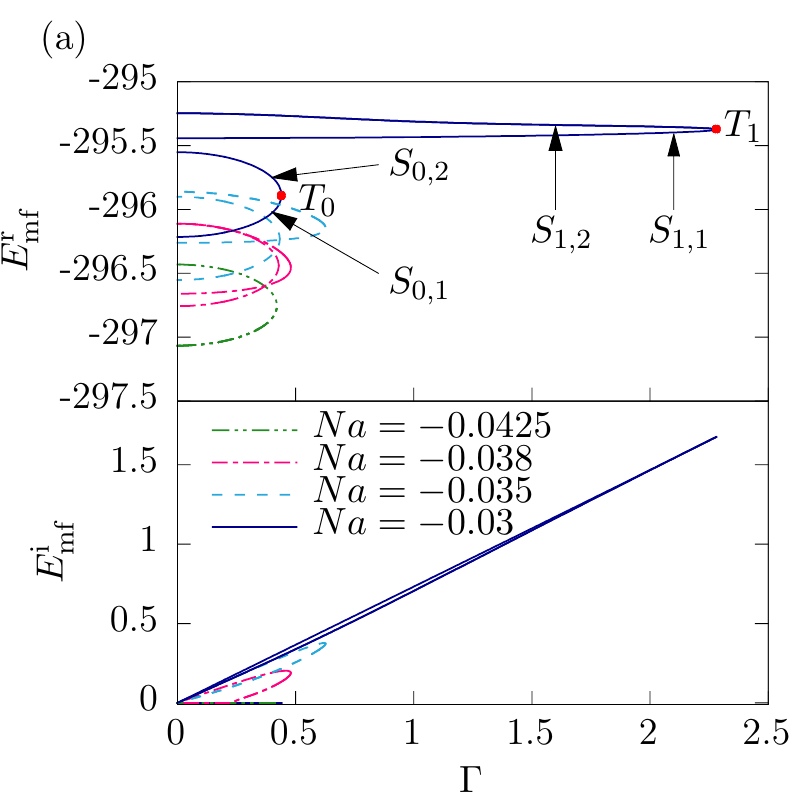}
  \includegraphics[width=\columnwidth]{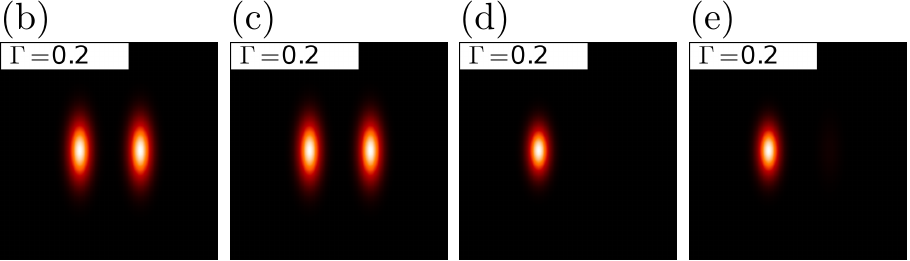}
  \caption{(Color online) (a) Mean-field energy of the stationary
    states in the repulsive configuration as a function of the
    gain-loss parameter $\Gamma$ for different values of $Na$. The
    upper panel shows the real and the lower one the imaginary part.
    The dipole strength is set to $Na\sub{dd}=0.3$. For $Na=-0.03$ the
    arrows in (a) point at the lines of the states for which in (b)
    $S_{0,1}$, (c) $S_{0,2}$, (d) $S_{1,1}$, and (e) $S_{1,2}$ the
    absorption images are shown at $\Gamma=0.2$. The field of view is
    $1\times 1$ in the units given in the text. In (a) the two tangent
    bifurcations $T_0$ and $T_1$ are labeled with red dots.}
  \label{fig:PT-rep-energy1}
\end{figure}
In \Fig~\ref{fig:PT-rep-energy1} results for $Na_{\mathrm{dd}}=0.3$
and several values of the scattering length $Na$ are given. We show
the mean-field energy as a function of the gain-loss parameter
$\Gamma$. For $Na=-0.03$ (solid blue line) at $\Gamma=0$ four
different states are present. This denotes the situation of a real
double-well potential and it is a remarkable result that the number of
states is larger in the dipolar system than in the non-dipolar one for
the corresponding situation, as shown in \Fig~\ref{fig:non-dipolar}
for $Na=-0.0044$ (double-dotted dashed light-blue line) at $\Gamma=0$.
Although this might stimulate attempts to gain a more complete picture
for the stationary states of non-dipolar and dipolar BEC in a real
double-well potential, we will concentrate on the investigation of the
effects introduced by gain and loss here. The two states with the
higher mean-field energy $S_{1,1}$ and $S_{1,2}$ break the \PT
symmetry of the external potential as can be seen in
\Figs~\ref{fig:PT-rep-energy1}(d) and (e). This leads to a finite
imaginary part of the mean-field energy for $\Gamma\neq 0$. By the
application of the \PT operator to these \PTs broken states
$S_{1,\alpha}$; $\alpha=1,2$ two new states $S_{1,\alpha}^\prime$
which have complex conjugate $E_\mathrm{mf}$ can be generated. These
new states $S_{1,\alpha}^\prime$ are also solutions of the
time-independent GPE. However, we omit the states
$S_{1,\alpha}^\prime$ in \Fig~\ref{fig:PT-rep-energy1}(a) and in the
following discussion as the analysis of these states is the same as
for the states $S_{1,\alpha}$. Note though that the dynamical behavior
is different. The lower states $S_{0,1}$ and $S_{0,2}$ preserve the
\PT symmetry and thus yield real $E_\mathrm{mf}$. Both pairs of states
$S_{0,\alpha}$ and $S_{1,\alpha}$ disappear in two separate tangent
bifurcations $T_0$ and $T_1$, respectively. Altogether six different
states are present in the energy range of \Fig~\ref{fig:non-dipolar}:
Two \PTs symmetric states, and two pairs of \PTs broken states.

\begin{figure}[tb]
  \centering
  \includegraphics[width=\columnwidth]{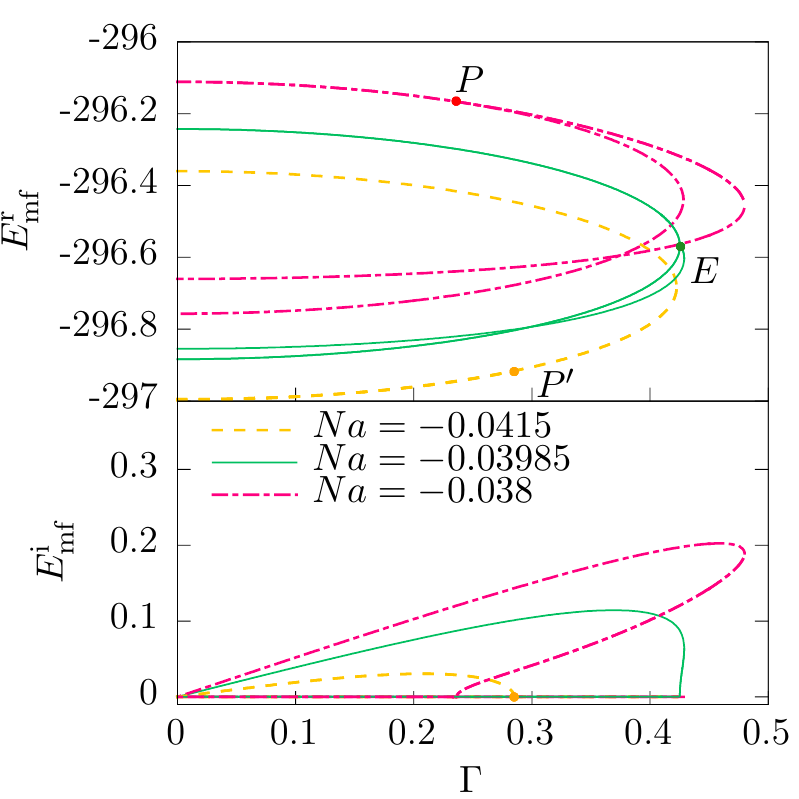}
  \caption{(Color online) Mean-field energy as a function of the
    gain-loss parameter $\Gamma$, where the upper panel shows the real
    and the lower one the imaginary part. The dipole strength is set
    to $Na\sub{dd}=0.3$. The points $P$ and $P^{\prime}$ denote
    pitchfork bifurcations for different values of the scattering
    length. At the point $E$ the pitchfork bifurcation has merged with
    the tangent bifurcation $T_0$ (see \Fig~\ref{fig:PT-rep-energy1})
    of the \PTs symmetric states. Note that for $Na=-0.0415$ the \PTs
    broken state for values of $\Gamma < P^\prime$ has almost the same
    real part of the energy as the \PTs symmetric one and lies on top
    of that in the upper panel.}
  \label{fig:PT-rep-energy2}
\end{figure}

Decreasing the scattering length $Na$ causes
$E_\mathrm{mf}^\mathrm{r}$ of the \PTs broken states to become smaller
and approach the values of the \PTs symmetric states. In this process
the mean-field energy of the state $S_{1,1}$ crosses both of the \PTs
symmetric states $S_{0,\alpha}$ (see e.g.\ dashed blue curve for
$Na=-0.035$ in \Fig~\ref{fig:PT-rep-energy1}(a)), yet the crossing
point is no exceptional point, i.e.\ the wave functions are diverse. A
further decrease of $Na$ has an effect similar to that observed in
\Refs~\cite{Dast2013,Dast2013jphysa}. From here on we include
\Fig~\ref{fig:PT-rep-energy2} for a detailed discussion. The state
$S_{1,2}$ separates in a pitchfork bifurcation from the \PTs symmetric
state $S_{0,2}$ (see red dashed-dotted line for $Na=-0.038$ in
\Fig~\ref{fig:PT-rep-energy2}).
The pitchfork bifurcation is labeled $P$ in
\Fig~\ref{fig:PT-rep-energy2}, whereas the tangent bifurcations of the
$S_{0,\alpha}$ and $S_{1,\alpha}$ states are denoted $T_0$ and $T_1$,
respectively, in \Fig~\ref{fig:PT-rep-energy1}. For the scattering
length $Na \approx -0.03985$ (solid green line in
\Fig~\ref{fig:PT-rep-energy2}) the tangent bifurcation $T_0$ of the
\PTs symmetric states and the pitchfork bifurcation $P$ merge in one
point, marked with a green dot, labeled $E$ in
\Fig~\ref{fig:PT-rep-energy2}(a). This behavior in a variation of the
scattering length $Na$ is therefore analogous to that in the
non-dipolar case. Yet, here the \PTs broken state vanishes for larger
$\Gamma$ in an additional tangent bifurcation $T_1$. It is a
remarkable fact that here the point $E$ appears at nontrivial
parameters and particularly for a finite nonlinearity and we will
revisit this point in the outlook. At $E$ the energy eigenvalues
obviously become real and the wave function preserves the \PT
symmetry. Note that the tangent bifurcation $T_1$ has not merged with
this point and remains separate at a slightly higher value of
$\Gamma$.

Decreasing the scattering length further leads $P$ to move down the
other \PTs symmetric state $S_{0,1}$, see e.g.\ the dashed yellow line
for $Na=-0.0415$ in the lower panel of \Fig~\ref{fig:PT-rep-energy2}.
There, the tangent bifurcation $T_1$ has already merged with the
pitchfork bifurcation so that the state $S_{1,1}$ has disappeared and
the state $S_{1,2}$ (see \Fig~\ref{fig:PT-rep-energy1}) is the only
remaining \PTs broken state. If the scattering length is tuned even
lower, $P$ is shifted to smaller values of $\Gamma$ until the
remaining \PTs broken state eventually disappears and only \PTs
symmetric states are left (see dashed-double-dotted green line for
$Na=-0.0425$ in \Fig~\ref{fig:PT-rep-energy1}). If we decrease $Na$
from there on, both \PTs symmetric solutions would disappear as well.
This behavior is well-known for a condensate in a real double-well
potential \cite{Milburn1997}.

\subsection{Stability and dynamics}
\label{sec:PT-rep-dynamics}

\begin{figure}[t]
  \centering
  \includegraphics[width=\columnwidth]{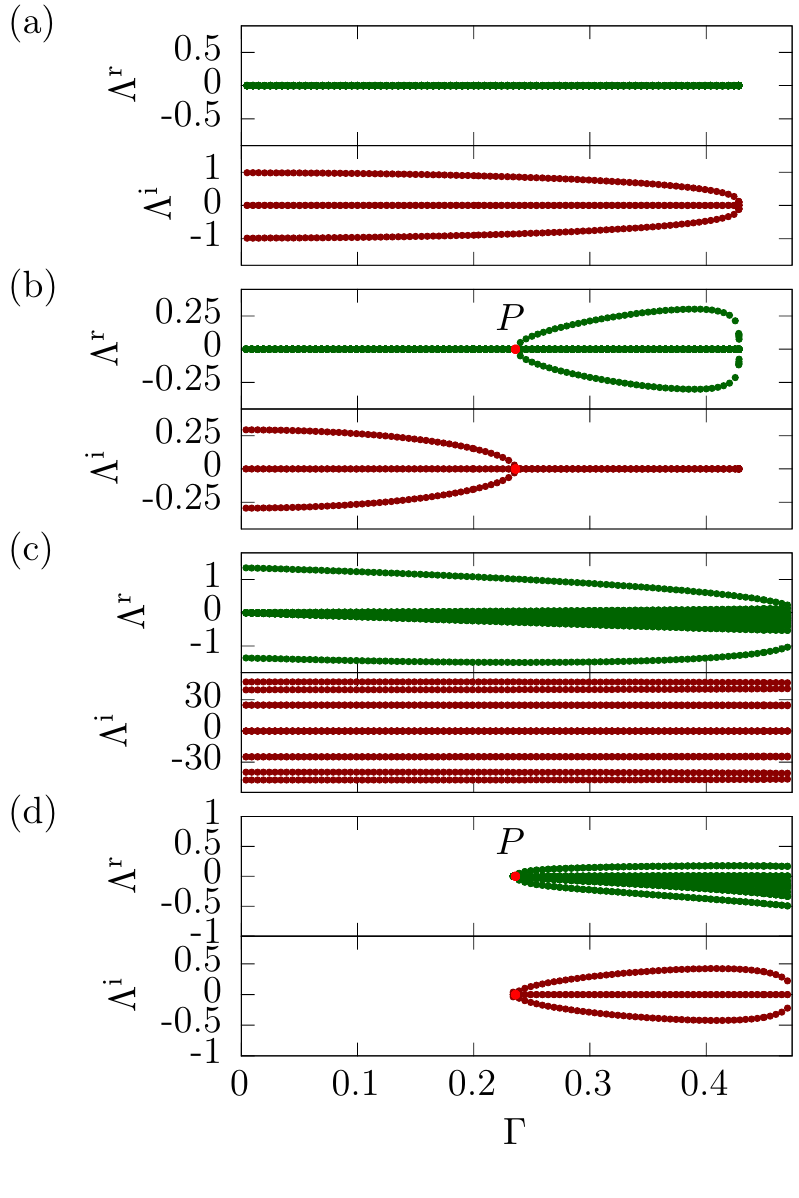}
  \caption{(Color online) Eigenvalues $\Lambda=\Lambda^\mathrm{r}+\ii
    \Lambda^\mathrm{i}$ of the Jacobian \eqref{eq:jacobian} as
    functions of the gain-loss parameter $\Gamma$ with $Na=-0.038$,
    and $Na\sub{dd}=0.3$. \Figs~(a)--(d) show the states $S_{0,1}$,
    $S_{0,2}$, $S_{1,1}$, and $S_{1,2}$, respectively. In (a) all
    eigenvalues are imaginary with vanishing real parts implying
    stable fixed points. The upper state, shown in (b) is stable up to
    the pitchfork bifurcation $P$. The \PTs broken states
    $S_{1,\alpha}$; $\alpha=1,2$ shown in (c) and (d) are unstable in
    the whole range of $\Gamma$. In (d) the pitchfork bifurcation $P$
    can be seen, where for values of $\Gamma < \Gamma_P$ only the
    state $S_{0,2}$ shown in (b) survives.}
  \label{fig:PT-rep-eigenvalues}
\end{figure}
The linear stability of the stationary points is investigated by the
use of the method presented in \Sec~\ref{sec:method}. The vanishing
real parts of all stability eigenvalues of the Jacobian
(\ref{eq:jacobian}) correspond to stable fixed points. In
\Fig~\ref{fig:PT-rep-eigenvalues} the stability eigenvalues for
$Na=-0.038$ (red dashed-dotted line in \Fig~\ref{fig:PT-rep-energy2})
are shown for all four states. The discussion of the states obtained
by the application of the \PTs operator is analogous to the one of the
states $S_{1,\alpha}$, except for the fact that in one case the norm
is increasing and for the other case decreasing for small periods in
time. The lowest-lying state $S_{0,1}$ is stable in the whole range of
$\Gamma$ as it is not involved in any pitchfork bifurcation with a
\PTs breaking state. This is different for the state $S_{0,2}$ (see
\Fig~\ref{fig:PT-rep-eigenvalues}(b)) which looses its stability
around the pitchfork bifurcation at $\Gamma_P \approx 0.24$. The \PTs
broken states shown in \Figs~\ref{fig:PT-rep-eigenvalues}(c),(d) are
unstable in the whole range, where they exist, as expected due to the
complex energy eigenvalues.

\begin{figure}[t]
  \centering
  \includegraphics[width=\columnwidth]{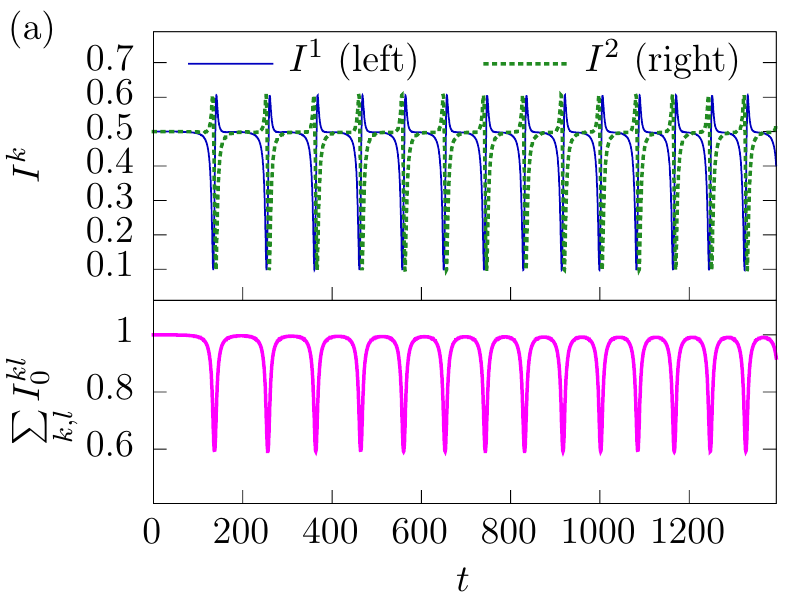}
  \includegraphics[width=\columnwidth]{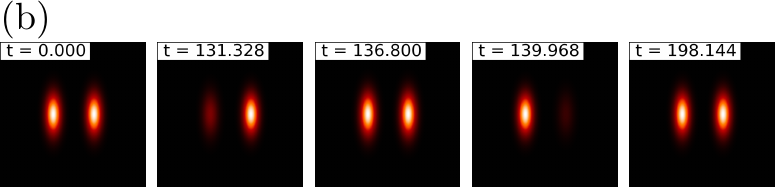}
  \caption{(Color online) Real-time evolution of the state $S_{0,2}$
    for $Na_{\mathrm{dd}}=0.3$, $Na=-0.038$, and $\Gamma=0.2$. No
    additional perturbances other than numerical fluctuations have
    been added to the initial state. (a) The upper panel shows the
    populations $I^k=\langle g^k | g^k \rangle$, $k=1,2$ of the left
    and the right well. The lower panel shows the overall norm of the
    wave function. In (b) absorption images during the first
    oscillation are plotted with the parameters given in
    \Fig~\ref{fig:PT-rep-energy1}. It can be seen that during the
    oscillation the wave function takes the shape of the different
    \PTs broken states.}
  \label{fig:PT-rep-dyn-osci-001}
\end{figure}
The real-time evolution of the stable state $S_{0,1}$ provides no
further insight as all parameters stay constant. Yet, the
corresponding evolution of the state $S_{0,2}$ reveals some very
interesting effects. In \Fig~\ref{fig:PT-rep-dyn-osci-001} the
real-time evolution of this \PTs symmetric state with a higher
mean-field energy than $S_{0,1}$ is shown. At first the state remains
approximately constant. Then a nonlinear oscillation between the wells
sets in. The shape of the wave function, which is illustrated by the
absorption images in \Fig~\ref{fig:PT-rep-dyn-osci-001}(b), passes
through states similar to the states shown in
\Figs~\ref{fig:PT-rep-energy1}(b)--(e). This behavior can be
interpreted in the following way. As $\Gamma < \Gamma_P$ the \PTs
broken state has not separated from the state $S_{0,2}$ and the linear
stability analysis declares the initial state stable. Yet, in this
complex system, where amongst others the loss of norm conservation
influences the dynamics, it is not sufficient to monitor the
eigenvalues of the linearized problem, but an investigation of the
full dynamics is required.

\begin{figure}[t]
  \centering
  \includegraphics[width=\columnwidth]{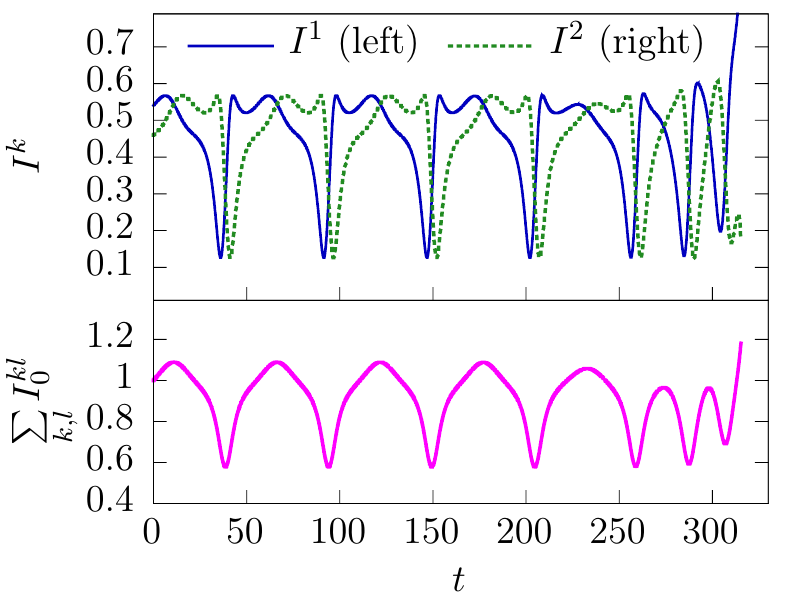}
  \caption{(Color online) Real-time evolution of the state $S_{1,2}$
    for $Na_{\mathrm{dd}}=0.3$ and $Na=-0.038$. In contrast to
    \Fig~\ref{fig:PT-rep-dyn-osci-001}, here $\Gamma=0.25$ is slightly
    above the value $\Gamma_P$ of the pitchfork bifurcation (see
    \Fig~\ref{fig:PT-rep-eigenvalues}(b)), i.e.\ in the unstable
    regime. Yet, the nonlinear coupling dynamically stabilizes the
    state for several oscillations until the condensates finally
    collapses.}
  \label{fig:PT-rep-dyn-osci-002-PT03_04}
\end{figure}
Interestingly, not only the amplitude shows that the oscillations
observed in \Fig~\ref{fig:PT-rep-dyn-osci-001} cannot be classified as
small. This is already observable in comparison with the smallest
finite eigenvalue $\abs{\Lambda^\ii\left( \Gamma=0.2
  \right)}_\mathrm{min} \approx 1$ as the corresponding time scale
would be one order of magnitude smaller than the time scale of the
oscillations in \Fig~\ref{fig:PT-rep-dyn-osci-001}. This confirms the
fact that a strong nonlinear coupling between the \PTs symmetric and
the \PTs broken states takes place. The strong coupling can already be
seen by a close look e.g.\ at the pitchfork bifurcation $P$. There is
a small gap between the values of $\Gamma_P$ obtained from the
energies and the stability eigenvalues, which has been shown to
originate from the nonlinearity in \Ref~\cite{Loehle2014}. The
importance of the influences of both the loss of norm conservation and
nonlinearity is impressively demonstrated in
\Fig~\ref{fig:PT-rep-dyn-osci-002-PT03_04}, where these prevent the
collapse of the condensate by a dynamical stabilization mechanism for
several oscillations. This is similar to the behavior of the
non-dipolar case described in \Ref~\cite{Haag14} by a projection on
the Bloch sphere, yet, this projection is not possible in the dipolar
system. More drastically, we found cases, where a state that is
unstable with respect to small perturbations, is dynamically
stabilized completely, regarding the collapse. The dynamical behavior
in such cases is similar to the nonlinear oscillations shown in
\Fig~\ref{fig:PT-rep-dyn-osci-001}. The larger the energetic distance
between the states is, the smaller the coupling gets. A further
increase of the gain-loss parameter suppresses the stabilization
mechanism and for $\Gamma=0.4$ the local collapse is induced already
by small perturbations of the stationary state.

\section{Conclusion}
\label{sec:conclusion}

We have shown that dipolar BECs in a \PTs symmetric double-well
potential feature real stationary solutions in parts of the parameter
space. Additionally, \PTs broken states are present, and in a distinct
range of the scattering length $Na$ one more pair of states is present
in the corresponding energy range than in the similar non-dipolar
system \cite{Dast2013}. The pair of states vanishes in a tangent
bifurcation together with the \PTs broken states observed also in
\cite{Dast2013}. Furthermore, in the dipolar system a point is found,
where the pitchfork bifurcation of the \PTs broken states merges with
the tangent bifurcation of the \PTs symmetric ones. This singular
point that has been shown in \Ref~\cite{Dast2013jphysa} to be related
to an exceptional point of order 4 is found for a nontrivial value
$Na\neq 0$ of the nonlinearity.

We found a strong influence of the nonlinearity on the dynamics of the
system. Often a linear stability analysis is not sufficient to
describe condensate wave functions close to stationary states. A
nonlinear coupling is capable of stabilizing states predicted to be
unstable by the linear stability eigenvalues or destabilizing states
which are supposed to be stable.

From the fact that we found similar results for the attractive
configuration one might ask the question if only the long-range nature
of the DDI causes the qualitative picture. Then, such results should
be observable e.g.\ with an isotropic long-range interaction as the
$1/r$-interaction \cite{ODell00a}. Furthermore, an appropriate matrix
model has to be developed to describe the additional states and
bifurcations. For a deeper understanding of the effects of the DDI the
system should be extended to a \PTs symmetric triple-well potential,
where it is easier to distinguish between on-site and long-ranged
effects. In the investigation of the singular point $E$ in
\Fig~\ref{fig:PT-rep-energy2} an analytic continuation and encircling
the point in the complex plane could reveal the properties of this
interesting feature. In particular, the encircling will state clearly
whether or not a true EP4 has been found.

The non-dipolar system of \Refs~\cite{Dast2013,Dast2013jphysa} can be
regarded as a subsystem of a generic Hermitian system -- in that case,
e.g.\ a four-well system \cite{Kreibich13a}. Thereby, the outer wells
serve as reservoirs of particles and constitute gain and loss,
proposing an experimental realization. However, with the DDI the
reservoirs are coupled to the inner wells by the long-ranged DDI and
thus the realization of a \PTs symmetric dipolar system requires a
different approach. Yet, this would be an interesting topic to
investigate as the correspondence to a larger system might allow for
the drawing of conclusions to the answers concerning the mechanisms in
large dipolar systems.

\section*{Acknowledgments}
R.F.\ is grateful for support from the
Landes\-graduierten\-f\"orderung of the Land Baden-W\"urttemberg. This
work was supported by Deutsche For\-schungs\-ge\-mein\-schaft.


%

\end{document}